\documentclass[%
 reprint,
 superscriptaddress,
 amsmath,amssymb,
 aps,
 pra,
longbibliography,
onecolumn
]{revtex4-1}

\usepackage{graphicx}% Include figure files
\usepackage{dcolumn}% Align table columns on decimal point
\usepackage{bm}% bold math
\begin{document}

\preprint{APS/123-QED}

\title{Asymmetric concentration dependence of segregation fluxes in granular flows}

\author{Ryan P. Jones}
\affiliation{Department of Mechanical Engineering, Northwestern University}

\author{Austin B. Isner}
\affiliation{Department of Chemical and Biological Engineering, Northwestern University}

\author{Hongyi Xiao}
\affiliation{Department of Mechanical Engineering, Northwestern University}

\author{Julio M. Ottino}
\affiliation{Department of Mechanical Engineering, Northwestern University}%Lines 
\affiliation{Department of Chemical and Biological Engineering, Northwestern University}
\affiliation{The Northwestern Institute on Complex Systems (NICO), Northwestern University}%Lines 

\author{Paul B. Umbanhowar}
\affiliation{Department of Mechanical Engineering, Northwestern University}

\author{Richard M. Lueptow}
\affiliation{Department of Mechanical Engineering, Northwestern University}
\affiliation{Department of Chemical and Biological Engineering, Northwestern University}
\affiliation{The Northwestern Institute on Complex Systems (NICO), Northwestern University}%Lines 

\date{\today}% It is always \today, today,

\begin{abstract}
 We characterize the local concentration dependence of segregation velocity and segregation flux in both size and density bidisperse gravity-driven free-surface granular flows as a function of the particle size ratio and density ratio, respectively, using discrete element method (DEM) simulations.  For a range of particle size ratios and inlet volume flow rates in size-bidisperse flows, the maximum segregation flux occurs at a small particle concentration less than 0.5, which decreases with increasing particle size ratio.  The segregation flux increases up to a size ratio of 2.4 but plateaus from there to a size ratio of 3.  In density bidisperse flows, the segregation flux is greatest at a heavy particle concentration less than 0.5 which decreases with increasing particle density ratio.  The segregation flux increases with increasing density ratio for the extent of density ratios studied, up to 10.  We further demonstrate that the simulation results for size driven segregation are in accord with the predictions of the kinetic sieving segregation model of \citet{Savage1988}.

\end{abstract}

\maketitle

\section{\label{sec:Intro}Introduction}

Flowing mixtures of granular material with differing properties, including size \citep{Chen2008, FanKinematics, Gray2005, Hill2014, Jing2017, Schlick2015AIChE, SchlickPoly}, density \citep{Jain2005, Khakkar1997, Liao2015, Saxton2008, Shinohara2003, Tripathi2013, Tunuguntla2017, XiaoDensity}, surface roughness \citep{Gillemot2017, Pohlman2006}, and shape \citep{Borzsonyi2013, Pereira2017}, tend to segregate, and they are common in geophysical flows \citep{Baker2016, Gray2010, Johnson2012, Woodhouse2012} and industrial settings such as during hopper filling and discharging \citep{Ketterhagen_Hopper, Standish_hopper, Shinohara__density_hopper}, in rotating tumblers \citep{Cleary2000, Jain2005, Khakhar1999, Maione2015, McClung2007, Schlick2015}, and in chute flow \citep{Marks2012, Savage1988, Thornton2012}.  The simplest explanation for segregation  relies on the idea, for size-disperse mixtures, that small particles fall through voids generated between large particles and accumulate in the lower regions of the flowing layer, while large particles are forced upward by concentrated regions of small particles.  Quantitative models of segregation have been developed and have now reached a state where accurate prediction is possible for a range of material properties and flow geometries \citep{Gray2018, Gray2006, Thornton2006, Tunuguntla2014, Schlick2015AIChE, SchlickPoly, Fan2014, Sam2018, Deng2018, Schlick2015}. 

Early experimental research on segregation in granular materials characterized segregation by tracking the center of mass of one of the species \citep{Harris1970, LawrenceBeddow1, LawrenceBeddow2, Scott1975, Williams1967, Williams1976} or by following tracer particles \citep{Campbell1973, Williams1976} finding that the effects of size differences on segregation are proportionally stronger than density differences, though both can result in significant segregation \citep{Williams1976}.  Of note are observations nearly 50 years ago by \citet{LawrenceBeddow1, LawrenceBeddow2} that segregation in flowing binary mixtures initially increases with increasing size ratio, $R_{S} = d_{l}/d_{s}$, where $d_{l}$ and $d_{s}$ are the large and small particle diameters, respectively, but plateaus in the interval $2.5 < R_{S} < 5$ and then decreases with further increases in $R_{S}$.  They also observed that segregation is greatest when the volume concentration of small particles, $c_{s}$, is between $15\%$ and $30\%$.  In this paper we consider both of these prescient observations in greater detail.

To explain the observed segregation behavior in granular mixtures, several models were proposed.  The screening mechanism of segregation by \citet{Shinohara1970} envisioned that large particles form a screen-like set of openings through which small particles percolate, a concept now referred to as ``kinetic sieving" \citep{Savage1988}.  This idea was extended by \citet{Williams1976} to include the rate of local particle rearrangement into configurations in which it is easier for small particles to fall between shear generated gaps.  Along similar lines, \citet{Cooke1979} proposed a statistical mechanics model for small particles falling into gaps between large particles.  A key aspect of this model is that the segregation rate depends linearly on the shear rate and exponentially on $R_{S}$.  

The ``kinetic sieving" mechanism and statistical mechanics model for dense granular flows of bidisperse mixtures were combined and expanded by \citet{Savage1988} to include, additionally, the probability of a particle falling downward under gravity into a shear generated void and the probability of ``squeeze expulsion" in which large and small particles are equally likely to be squeezed upward by particles below them.  The model expresses the local motion of particles species $i$ normal to the free surface as a segregation, or percolation, velocity, $w_{p,i}$:
\begin{equation}
 \label{segVel}
 w_{p,i}=w_{i}-w,
\end{equation}
where $w_{i}$ is the species velocity and $w$ is the local bulk velocity averaged over all species, both normal to the free surface. This model (the ``Savage and Lun" model) is discussed in detail in Section \ref{sec:SavageLun}, but here we note that in the model $w_{p,i}$ depends on the particle size ratio, the volume concentration of particles of the other species, or ``concentration complement" $1-c_{i}$, the local shear rate, $\dot{\gamma}$, local void properties, the flowing layer thickness, $\delta$, and other parameters \citep{Savage1988}.  A simplified approximation of the Savage and Lun model can be expressed as
\begin{equation} 
 \label{simpleSL} 
 w_{p,i}=\dot{\gamma}d_{s}f(c_{i},R_{S})(1-c_{i}),
\end{equation}
where $f(c_{i},R_{S})$ incorporates the percolation velocity's nonlinear dependence on particle concentration and particle size ratio.  

Other first order expressions for the segregation velocity have been proposed, though all retain the linear dependence on the local particle concentration complement, $1-c_{i}$.  \citet{Gray2005} proposed a segregation velocity with explicit dependence on gravity, $g$, and the repose angle of the free surface, $\alpha$: 
\begin{equation}
 \label{GrayEquation}
 w_{p,i} = C_{0}g\cos(\alpha)(1-c_{i}),
\end{equation}
where $C_{0}$ is a coefficient related to inter-particle drag.  \citet{Hill2014} proposed a stress-based segregation velocity model of the form 
\begin{equation}
 \label{HillEquation}
 w_{p,i} = C_{1}\dfrac{dP}{dz}(1-c_{i}),
\end{equation}
where $C_{1}$ describes the stress partitioning with a linear drag coefficient and $P$ is the pressure.  \citet{Fan2014} simplified the Savage and Lun model to    
\begin{equation} 
 \label{FanEquation} 
 w_{p,i}=\dot{\gamma}S(1-c_{i}),
\end{equation}
where, $S = d_{s}f(R_{S})$, called the segregation coefficient, is an empirical parameter dependent on the particle size ratio for size bidisperse flows of spherical, mm-sized particles \citep{Schlick2015AIChE}.  \citet{XiaoDensity} demonstrated that the same model accurately describes segregation in flows of density bidisperse spherical particles having the same radius, but where $S$ is now a function of the density ratio, $R_{D} = \rho_{h}/\rho_{l}$, where $\rho_{h}$ and $ \rho_{l}$ are the densities of the heavier and lighter particles, respectively.

For bidisperse granular materials with constant volume fraction segregating normal to the free surface, mass conservation requires that 
\begin{equation}
 \label{massConservation}
 \Phi_{up}=-\Phi_{down}, 
\end{equation}
where $\Phi_{up}$ and $\Phi_{down}$ are the segregation volume fluxes of the upward segregating species and downward segregating species, respectively.  The segregation flux for species $i$, $\Phi_{i}$ is 
\begin{equation}
 \label{simpleFlux}
 \Phi_{i}=w_{p,i}c_{i}.
\end{equation}
Because in binary mixtures with constant volume fraction the species concentration $c_{up}=1-c_{down},$ Eq.~\ref{simpleFlux} implies that $w_{p,up} = -w_{p,down}$ only when $c_{up} = c_{down} = 0.5$; in general, the segregation velocities are not equal.
 
A consequence of assuming a segregation velocity linear in concentration like that in the models mentioned above (Eqs.~\ref{GrayEquation}-\ref{FanEquation}) is that the segregation flux is maximum at and symmetric about $c_{i} =  0.5$ for bidisperse mixtures.  However, recent experiments examining slowly sheared size bidisperse granular material in a confined annular shear cell with steady shear \citep{GollickAndDaniels} and in a shear cell with periodic shear \citep{VaartAndGray} have shown that the maximum segregation flux for a given size ratio occurs at concentrations of small particles $c_{s}<0.5$, confirming Lawrence and Beddow's 1969 observations \citep{LawrenceBeddow2}.  The annular shear cell experiments \citep{GollickAndDaniels} also show that the segregation rate does not increase monotonically with $R_{S}$, as predicted by the \citeauthor{Cooke1979} model \citep{Cooke1979}.

To address the observed asymmetry of the segregation flux with respect to $c_{s} = 0.5$, \citet{GajjarAndGray} proposed a two parameter cubic form of the segregation flux, 
\begin{equation}
 \label{grayForm}
 \Phi_{s} = \beta c_{s}(1-c_{s})(1-\kappa c_{s}),
\end{equation}
where $\beta$ is a magnitude coefficient and $\kappa$ is an asymmetry coefficient.  This expression yields a quadratic segregation velocity dependence on species concentration rather than the linear dependence in Eqs.~\ref{GrayEquation}-\ref{FanEquation}.  

In this study we characterize concentration dependent asymmetry in the segregation velocity in gravity-driven free-surface flows of bidisperse granular material over a range of bidispersities.  The results of this study can be used to improve continuum model predictions of segregation, which, although not the focus of this paper, is demonstrated in the supplemental material.  The remainder of the paper is organized as follows.  In Section \ref{sec:SizeSeg}, size driven segregation is investigated.  In Section \ref{sec:SavageLun} predictions of the Savage and Lun model \citep{Savage1988} are compared with size segregation results from DEM simulations, demonstrating this  model's ability to capture features observed in DEM results.    Section \ref{sec:Density} demonstrates that density-driven segregation exhibits asymmetry in the segregation flux with respect to concentration that is nearly identical with that found for size-driven segregation, which raises interesting questions about the size-based Savage and Lun model.  Section \ref{sec:Conclusion} presents our conclusions.

\section{\label{sec:SizeSeg}Segregation in size bidisperse mixtures}

\begin{figure}

	\center	
		\includegraphics[scale = .75,trim={0.5in 3.45in 0.5in 3.6in}, clip]{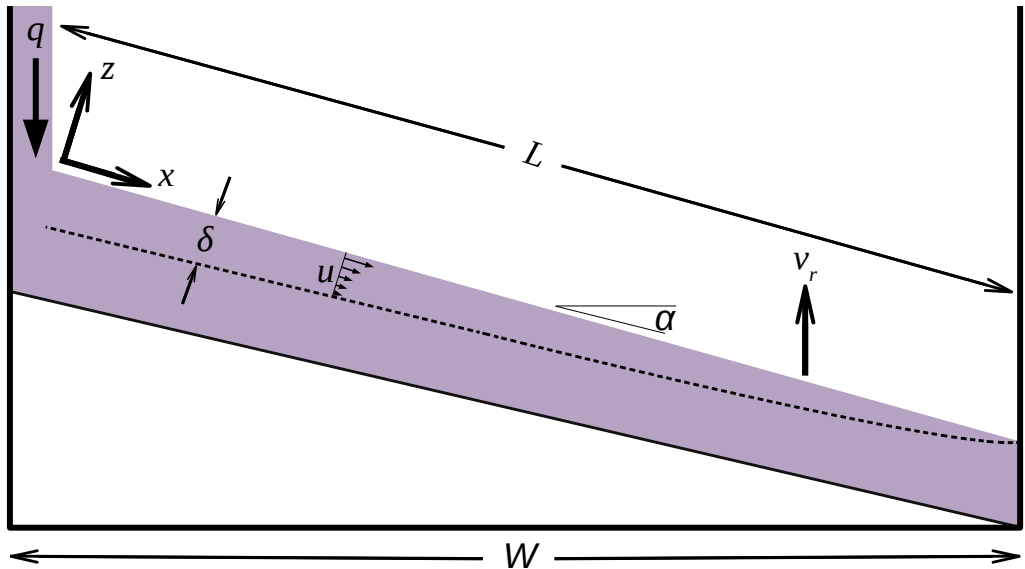} 
	\caption{Schematic of the quasi-2D bounded heap.  Particles fall onto the $W$ wide heap at a 2D volume flow rate, $q$, resulting in a free surface rise velocity $v_{r}$. The sloped lower boundary reduces the number of particles required for the DEM simulations.  The flowing layer (above dotted line) has thickness $\delta$ and length $L$.  Particles are continuously deposited from the bottom of the flowing layer onto the heap.  The coordinate system is rotated by the angle of repose, $\alpha$, and rises at $v_{r}$.  The thickness of the flowing layer is exaggerated and its shape is idealized.}   \label{Heap}
\end{figure}

Discrete element method (DEM) simulations of granular flows have reached the point where their results are nearly equivalent to those measured in corresponding experiment \citep{FanKinematics, Schlick2015AIChE, XiaoDensity}, but with the significant benefit that all the properties of the simulated flow are easily measured.  Accordingly, we use DEM simulation to study segregation in the context of a gravity-driven free surface flow of granular material in a one-sided quasi-2D bounded heap with a sloped lower boundary (to reduce the number of simulated particles), see Fig.~\ref{Heap}.  In the one-sided quasi-2D bounded heap geometry, particles fall onto the left side of the heap and flow down the heap in a thin flowing layer of thickness, $\delta$, at an angle of repose, $\alpha$.  The origin of the coordinate system, with streamwise coordinate, $x$, and surface normal coordinate, $z$, is coincident with the free surface and rises at the heap rise velocity, $v_{r}$. 

\subsection{\label{sec:DEM}DEM simulation methodology}

In all DEM simulations the heap container has a spanwise thickness $T/d_{l} = 6$ and length $W/d_{l} = 200$.  At the end of a typical simulation, the heap consists of approximately $10^{6}$ particles.  Spherical particles are fed by gravity at 2D volume flow rates, $q = \dot{m}/\rho T$, of 20\,cm$^{2}$/s or 40\,cm$^{2}/$s, where $\dot{m}$ is the mass flow rate and $\rho$ is the particle density, $\rho = 2500$\,kg/m$^{3}$.  The heap forms with an angle of repose, $\alpha$, slightly greater than the slope of the bottom wall.  Particles that touch the bottom wall become stuck to the wall, thus creating a rough bottom boundary.  Other boundaries are modeled as smooth walls.  A steady state is reached after the heap becomes sufficiently deep, typically about $\sim 10d_{l}$, to minimize kinematic effects from the bottom boundary \citep{FanKinematics}.  For size bidisperse simulations, particle diameters for each species are uniformly distributed with mean diameter, $d$, between $0.9d$ and $1.1d$ to reduce ordered packing.  Mean particle diameters range from 2 to 6\,mm to obtain size ratios from $1 \le R_{S} \le 3$.  

Standard DEM methods \citep{springDashpot, Schafer1996, Silbert07, Ristow2000} are used as described in detail previously \citep{FanKinematics, Schlick2015AIChE, XiaoDensity}.  Our in-house parallelized DEM code runs on an NVIDIA GTX 980 GPU or an NVIDIA GTX Titan X GPU installed in a workstation computer running Ubuntu 14.04 LTS and has been previously validated against experimental data for mm-sized glass particles in bounded heap flows of size bidisperse particles \citep{FanKinematics} and for mm-sized glass, ceramic, and steel particles in density bidisperse flows \citep{XiaoDensity}.  For all simulations a binary collision time of $t_{c}=10^{-3}$\,s and a restitution coefficient $e = 0.8$ are used as in previous simulations \citep{Schlick2015AIChE, FanKinematics}.  Particle-particle and particle-wall contacts both use a friction coefficient of 0.4.  The integration time step of $t_{c}/40=2.5 \times 10^{-5}$\,s ensures numerical stability for these flows \citep{Ristow2000}.  Simulation data is collected once the rise velocity is constant and spatially uniform.  

The segregation velocity and species concentration are calculated from spatial and temporal averages of the DEM simulation output.  Although other coarse-graining methods exist \citep{Weinhart2013}, the method described below is used similar to previous work \citep{Schlick2015AIChE}.  At each output time step, the particle data is binned into quadrilateral bins oriented along the free surface, with a stream-wise length of $3 d_{l}$, a height (normal to the free surface) of $1.25 d_{l}$, and a width equal to the spanwise extent of the heap.  The bins move upward with the free surface at $v_{r}$.  For averaging purposes, the partial volumes of particles overlapping bin boundaries are applied to the appropriate bin.  The species concentration in each bin is defined as 
\begin{equation}
  \label{concentrationCalc}
  c_{i}=\dfrac{\sum V_{i,j}}{V_{bin}},  
\end{equation}
where $V_{i,j}$ is the volume of each particle, $j$, of species $i$ and $V_{bin}$ is the total volume of all particles in the bin.  Although the sidewall friction can alter the flowing layer thickness, segregation is a local quantity driven by the local shear rate and local concentration. Since the spanwise variation of the velocity and concentration is small in narrow, quasi-2D heaps \citep{FanKinematics}, the spanwise average is used, consistent with previous studies \citep{Schlick2015,SchlickPoly,Fan2017}.  Furthermore, we have confirmed that the segregation velocity does not depend on the particle location relative to the sidewalls by comparing the segregation velocity measured near the walls to that measured midway between the sidewalls. The mean velocity of the $i^{th}$ species in the bin, $\mathbf{u_{i}}$, is based on the volume weighted velocity as
\begin{equation}
  \label{normalVelocity}
  \mathbf{u_{i}}=\dfrac{\sum \mathbf{u_{i,j}}V_{i,j}}{\sum V_{i,j}},  
\end{equation}
where $\mathbf{u_{i,j}}$ is the velocity of each particle, $j$, of species $i$.  The concentration and velocity values of each bin are then temporally averaged across the all output timesteps, which are separated by 0.05 seconds (2000 simulation timesteps).  The shear rate in a bin
\begin{equation}
  \label{shearCalc}
  \dot{\gamma}=\dfrac{du}{dz},  
\end{equation}
is calculated from the averaged bin velocities as a backward finite difference (the change in the streamwise velocity, $u$, between the target bin and the bin below it divided by the vertical separation) but is insensitive to the finite differencing method that is used.  The local segregation velocity, $w_{p,i}$, for each particle species in each bin within the flowing layer at each output time step is calculated using Eq.~\ref{segVel}.  

\subsection{\label{sec:DEMResults}Segregation velocity and flux}

\begin{figure}
	\center	
		\includegraphics[scale = 1]{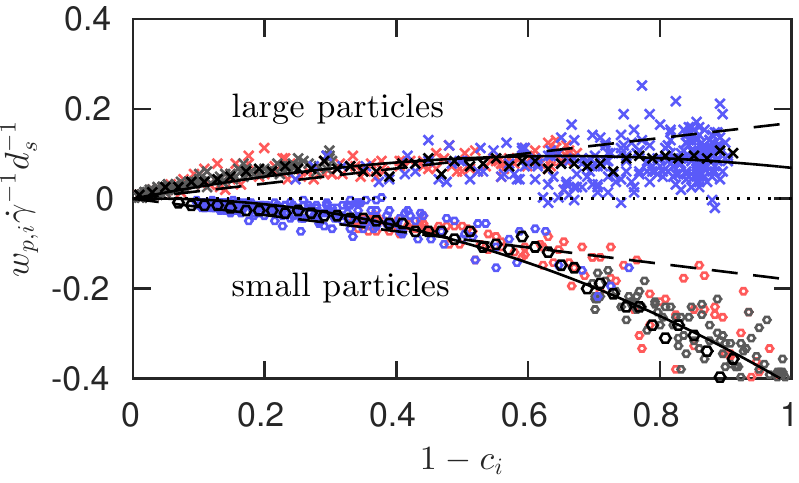} 
	\caption{Segregation velocity dependence on the local concentration complement, $1-c_{i}$, for $R_{S}=2.2$, $d_{s} = 2$\,mm, and $q = 40$\,cm$^{2}/$s.  Segregation velocity from simulations with particle inlet ratios $c_{s,inlet}=0.2$ (blue data points), $0.5$ (red data points), and $0.8$ (gray data points) and black data points are averaged over 0.02 wide concentration increments, x = large particles and o = small particles.  Dashed (- -) line is a linear fit as in previous work (Eq.~\ref{FanEquation}) \citep{Schlick2015AIChE} using only data from $c_{s,inlet}=0.5$, solid (--) curve is a quadratic fit using all data in the plot (Eq.~\ref{WPoly}). }\label{VelPoly}
\end{figure}

An example of the segregation velocity data for large and small particles is plotted in Fig.~\ref{VelPoly} for $R_{S} = 2.2$, $d_{s} = 2$\,mm, and $q = 40$\,cm$^2$/s.  Colored data points represent the local segregation velocity for different inlet concentrations calculated in each bin throughout the entire flowing layer averaged over the total number of output time steps.  Black data points are averaged over 0.02 increments of $1-c_{i}$ to minimize the scatter due to collisional diffusion and more clearly show the data trend.  The segregation velocity, $w_{p,i}$, is non-dimensionalized by $d_{s}$ and $\dot{\gamma}$ so the results can be considered in the context of Eqs.~\ref{simpleSL} and \ref{FanEquation}.  Note that previous studies evaluating the segregation velocity in the heap geometry considered 50:50 mixtures of small and large particles, corresponding to, $c_{s,inlet} = 0.5$ \citep{Schlick2015AIChE, Fan2014}.  However, under these conditions the small particle concentration is mostly in the range $0.3 < c_{s} < 0.7$.  The full range of concentrations $(0<c_{s}<1)$ are not observed because in the bounded heap small particles quickly percolate to the bottom of the flowing layer and are deposited onto the upstream portion of the heap while large particles are advected toward the downstream end wall.  The concentration of small particles in the flowing layer therefore continually decreases in the stream-wise direction until the flow reaches the bounding wall, limiting the lower limit of the small particle concentration so that data outside the range of concentrations mentioned above are never obtained.  To evaluate the segregation velocity over a wider range of concentrations, we conduct simulations with $c_{s,inlet}=0.2$, $0.5$, and $0.8$.  Note in Fig.~\ref{VelPoly} that $c_{s,inlet}$ does not affect the segregation velocity relation, it only affects the range of local concentrations within the flowing layer in the bounded heap geometry.  

Figure \ref{VelPoly} shows that large particles have a positive (upward) segregation velocity, and small particles have a negative (downward) segregation velocity.  The segregation velocity for species $i$ generally increases in magnitude with increasing concentration of the other species, $1-c_{i}$, though the dependence on $1-c_{i}$ is more dramatic for small particles.  A small particle among mostly large particles segregates faster than a small particle among mostly small particles.  Note that data points from different mixture inlet concentrations (different colors) occupy different portions of the concentration complement data, but also overlap, confirming that $c_{s,inlet}$ does not affect $w_{p,i}/\dot{\gamma}$, just its range in a bounded heap flow.  The data presented in Fig.~\ref{VelPoly} spans a broad range of shear rates, $0.7 < \dot{\gamma} < 23.0$ s$^{-1}$, due to the decreasing 2D volume flow rate, $q$, with streamwise position in the bounded heap geometry.  The wide range of percolation velocities for small particles as $1-c_{i}$ approaches 1 is expected when a very low number of small particles (one or two) are in a bin otherwise filled with large particles.  In this single particle limit, the percolation velocities are expected to vary widely due to random particle collisions and the probabilistic mechanics of kinetic sieving \citep{Savage1988}.  The segregation velocity does not show a significant dependence on the concentration gradient.  It should be noted that diffusion scales with the shear rate and particle diameter, $D\sim \dot{\gamma}d^{2}$ \citep{Fan2014, Utter2004}.  The diffusive flux is relatively small compared to the segregation flux, $D\partial c_{i} / \partial z<0.1 \Phi_{i}$.     

\begin{figure}
	\center	
		\includegraphics[scale = 1]{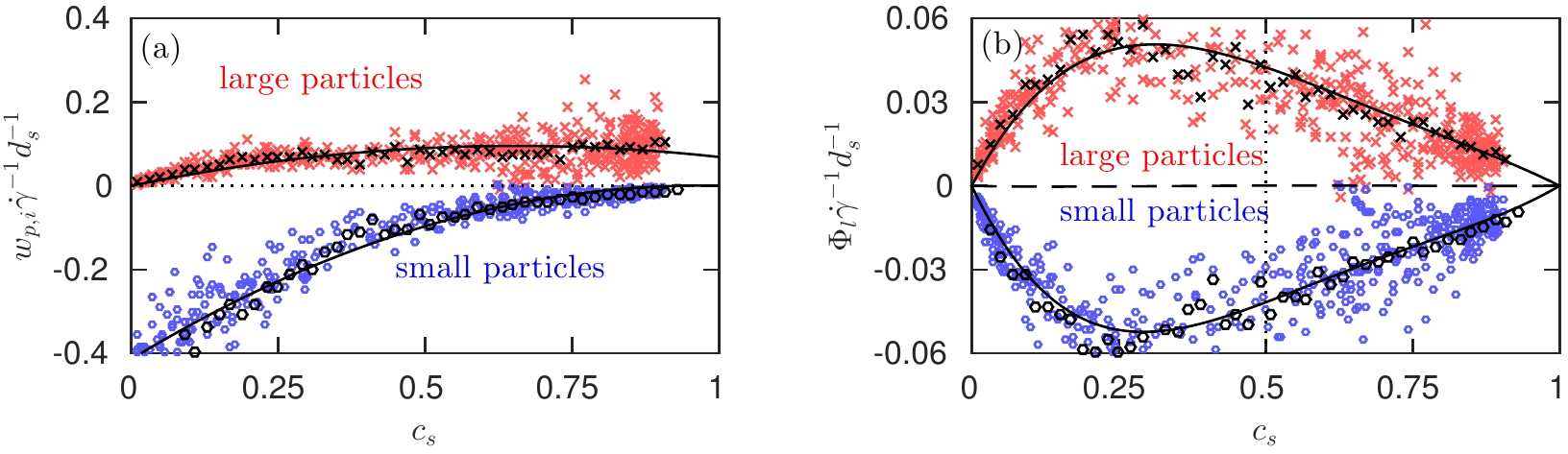}
	\caption{Segregation velocity and flux dependence on local small particle concentration, $c_{s}$, for $R_{S}=2.2$, $d_{s} = 2$\,mm, and $q = 40$\,cm$^{2}/$s.  (a) Segregation velocity with $c_{s,inlet}=0.2$, $0.5$, and $0.8$ (red x = large  particles, blue o = small particles; black data points are averaged over 0.02 wide increments of $1-c_{i}$).  Solid (--) curve is a quadratic fit using all data in plot (Eq.~\ref{WPoly}).  (b) Segregation flux with solid (--) curve from fits in (a).  Dashed (- -) line is the measured sum of the upward and downward fluxes, demonstrating conservation of mass.  Dotted vertical reference line at $c_{s} = 0.5$ highlights the asymmetry of flux with respect to concentration.}\label{FluxPoly}
\end{figure}

The same data can be plotted vs.\ the concentration of a single species, such as $c_{s}$ in Fig.~\ref{FluxPoly}(a).  Plotted in this way, it is evident that large particles and small particles rise or sink at different velocities at the same small particle concentration.  For instance, at $c_{s} = 0.3$ the magnitude of $w_{p,s}$ is about twice the magnitude of $w_{p,l}$. Thus, a small particle among many large particles sinks faster than surrounding large particles rise.  Likewise, at $c_{s} = 0.8$, the magnitude of $w_{p,s}$ for the small particles is small compared to $w_{p,l}$ for the large particles.  Thus, a large particle among many small particles rises faster than the small particles sink.  This behavior is expected and is why the concentration complement, $1-c_{i}$, is used in Eqs.~\ref{simpleSL}-\ref{FanEquation}.  What is more interesting is that the small particles can sink at a maximum percolation velocity as much as four times that of the maximum percolation velocity that a large particle rises.  In fact, for $c_{s} \le 0.5$ small particles sink much faster than large particles rise.

Despite the large differences in percolation velocity, mass is conserved, as demonstrated by plotting the fluxes of the two types of particles in Fig.~\ref{FluxPoly}(b).  At any small particle concentration, $c_{s}$, the flux of large particles upward equals the flux of small particles downward.  For the volume flux of small particles sinking downward to match that of large particles moving  upward, either more small particles need to be moving downward than large particles moving upward or the small particles must sink downward faster than the same number of large particles moving upward.  At low small particle concentrations, the small particles sink much faster than the large particles rise to conserve mass.  The consequence is an asymmetry in the segregation flux, which is maximum near $c_{s,peak} = 0.35$ as shown in Fig.~\ref{FluxPoly}(b).  Thus, the greatest local segregation flux occurs when the concentration of large particles is greater than the concentration of small particles, consistent with previous results \citep{LawrenceBeddow2, VaartAndGray, GollickAndDaniels}.  The difference between this study and these previous studies is that here we quantify this effect for a range of particle size ratios, $R_{S}$, of particles segregating in a flowing layer having a wide range of flow conditions along the length and depth of the flowing layer as well as representing the full range of relative concentrations of small and large particles rather than under limited flow and concentration conditions.  Since segregation is a local effect dependent on concentration, which varies throughout the flowing layer in the bounded heap flow, the segregation flux also varies throughout the flowing layer.  It is small in regions dominated by large or small particles ($c_{s}$ near 0 or 1) and largest for $c_{s} \approx 0.35$, for the conditions used to generate Fig.~\ref{FluxPoly}(b).

Returning to Fig.~\ref{VelPoly}, the data were fit (MATLAB Linear Least Squares function with representative data weighting and the bisquare outlier weighting) to both Eq.~\ref{FanEquation} (linear) and a quadratic polynomial.  Only data for $c_{s,inlet} = 0.5$ were used for the linear fit, consistent with previous work \citep{Schlick2015AIChE}.  Data for $c_{s,inlet} = 0.2,\ 0.5,$ and $0.8$ were used for the quadratic fit.  The linear fits (dashed) do not match the segregation velocity well for $1-c_{i} > 0.6$.  Using data for $c_{s,inlet} = 0.2,\ 0.5,$ and $0.8$ alters the linear fit slightly (not shown), but still does not capture the observed curvature of the segregation velocity data, evident in the concentration-averaged data (black).  Using the simplest non-linear curve, a quadratic fit (solid curves) better describe $w_{p,i}$, especially the downward curvature of the segregation velocity for both species when $1-c_{i} > 0.6$. Further, to capture the asymmetric segregation flux a non-linear segregation velocity is required.  Based on these results, we replace the expression in Eq.~\ref{FanEquation} for $w_{p,i}$ with
\begin{equation}
 \label{WPoly}
 w_{p,i}=d_{s}\dot{\gamma}\left[A_{R,i}+B_{R,i}(1-c_{i})\right](1-c_{i}),
\end{equation}
where $A_{R,i}$ and $B_{R,i}$ are fit coefficients that depend on the size ratio, $R_{S}$, and the species, $i$.  

When the segregation velocity relation is quadratic in $c_{i}$, the segregation flux (Eq.~\ref{simpleFlux}) is cubic in $c_{i}$: 
\begin{equation}
 \label{cubicFlux}
 \Phi_{i} = d_{s}\dot{\gamma}c_{i}[A_{R,i}(1-c_{i})+B_{R,i}(1-c_{i})^{2}].
\end{equation}
This form is identical to that proposed by \citet{GajjarAndGray} (Eq.~\ref{grayForm}) with the magnitude coefficient $\beta = A_{R,i}d_{S}\dot{\gamma}$ and the asymmetry coefficient $\kappa = -B_{R,i}/A_{R,i}$.  The form for the segregation flux suggested by \citet{GajjarAndGray} was based on results from experiments in an oscillatory shear cell for a single size ratio, $R_{S}=2.0$ \citep{VaartAndGray}.  

The segregation flux, shown in Fig.~\ref{FluxPoly}(b), demonstrates that even though the values for $A_{R,i}$ and $B_{R,i}$ are found independently for each species $i$ from the segregation velocity in Fig.~\ref{VelPoly}, the resulting segregation fluxes (solid curves) match the data very well.  Furthermore, the small and large particle segregation fluxes sum to zero (dashed curve) as they should based on mass concentration in an incompressible flow.  Thus, the fit to the DEM data provides a physical basis for the quadratic form for the segregation velocity (Eq.~\ref{WPoly}) and the cubic form of the segregation flux (Eqs.~\ref{grayForm} and \ref{cubicFlux}).
 	
\begin{figure*}
 		\centering
 		\includegraphics[scale = 1]{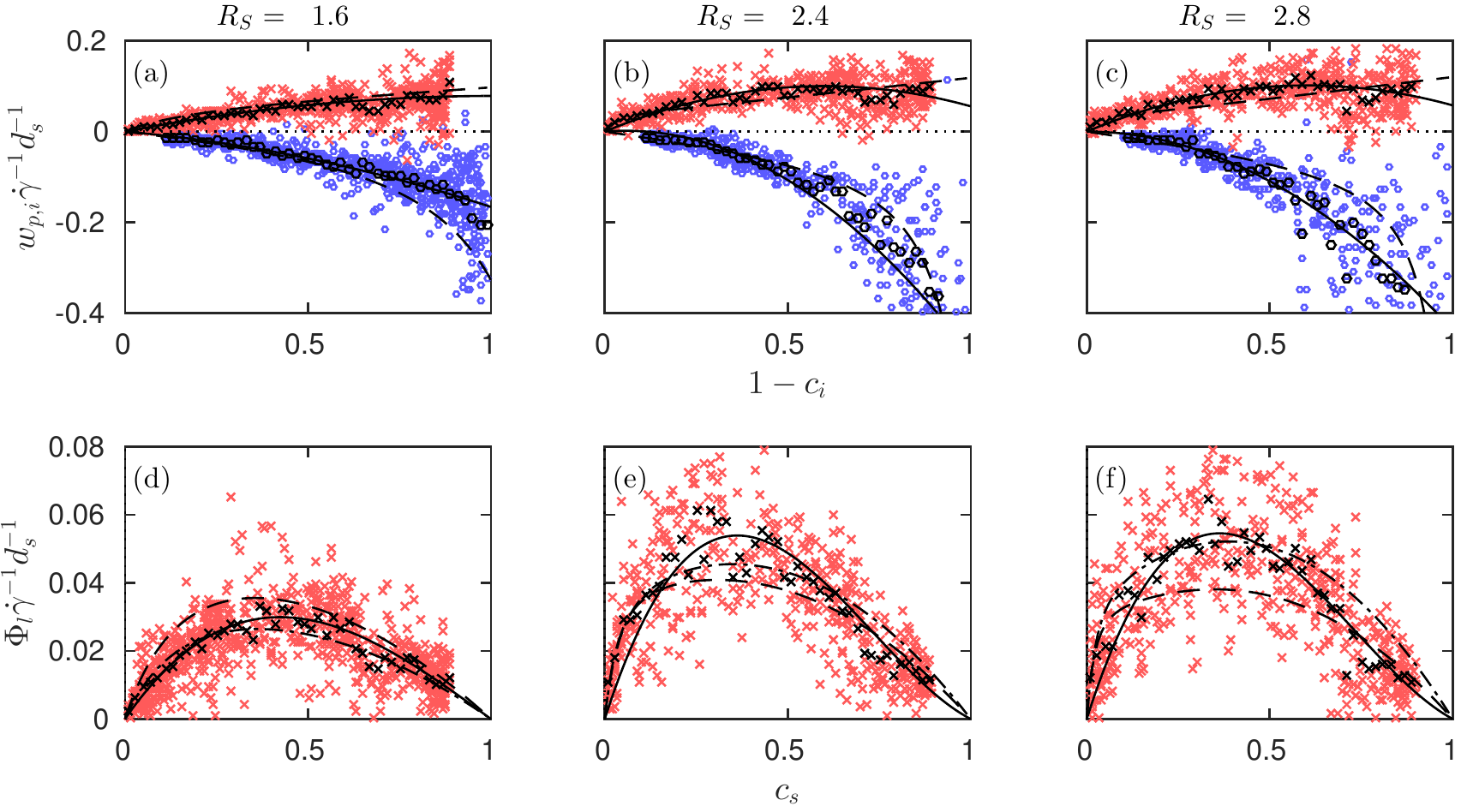}
 		\caption{(a-c) DEM segregation velocity data for large (red x) and small (blue o) particles vs.\ $1-c_{i}$ at various size ratios.  Black datapoints show mean segregation velocity in $0.02$ wide $1-c_{i}$ increments.  Solid (--) curves are the quadratic fit (Eq.~\ref{WPoly}) of the segregation velocity, dashed (- -) curves are the theoretical predictions of the SL-model for each particle species.  (d-f) DEM data for large particle segregation flux (red) and averaged over 0.02 wide concentration increments (black).  Solid (--) curves are the segregation flux based on the fit of the segregation velocity data (Eq.~\ref{cubicFlux}), dashed (- -) curves are the theoretical predictions of the SL-model (fit to all DEM data), dotted (- $\cdot$ -) curves are the theoretical predictions of the SL-model fit to only the DEM data for that $R_{S}$.  SL-model parameters were fit to DEM segregation velocity data for all data for $1.1 \le R_{S} \le 3.0$, resulting in parameter values  $\bar{E}=0.477$, $M/N=0.781$, $k_{\scriptscriptstyle{\text{av}}}=0.466$, $k_{\scriptscriptstyle{\text{LT}}}=1$, SL-model parameters were fit to DEM segregation velocity data for $R_{S} = 1.6, 2.4,$ or $2.8$, resulting in parameter values $\bar{E}=0.487, 0.517,$ and $ 0.532$, $M/N=0.5, 0.693,$ and $ 0.812$, $k_{\scriptscriptstyle{\text{av}}}=0.466, 0.466,$ and $ 0.466$, $k_{\scriptscriptstyle{\text{LT}}}=1, 1,$ and $ 1$, respectively.}
 		\label{fig:SL2}
 	\end{figure*}

To test whether Eqs.~\ref{WPoly} and \ref{cubicFlux} are valid for other particle size ratios, we conducted DEM simulations for $1.1 \leq R_{S} \leq 3.0$.  Both the quadratic fit for the segregation velocity and the resulting cubic fit for the segregation flux for large particles (solid curves) are shown for three additional size ratios, $R_{S}=1.6,\ 2.4,$ and $2.8$, in Fig.~\ref{fig:SL2}.  The segregation velocity and segregation flux data are again averaged in $0.02$ concentration increments and these average data points overlay the raw data.  For all size ratios, the quadratic fit (for segregation velocity) and resulting cubic curve (for flux) match the DEM simulation data well, as highlighted by the concentration-averaged data.  The figure also includes dashed curves demonstrating the fit of Savage and Lun's segregation model, discussed in more detail in Section \ref{sec:SavageLun}.  

\begin{figure}
	\center	
		\includegraphics[clip, trim=0cm 0cm 12.5cm 23cm,scale = 1.]{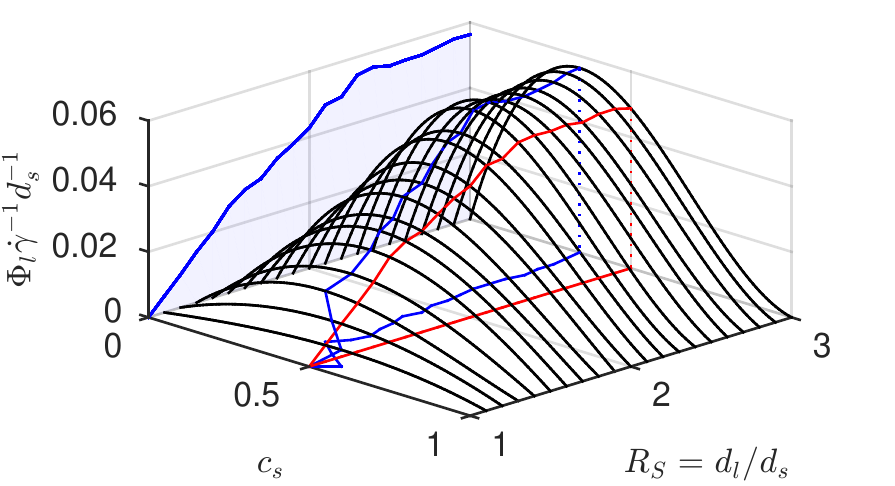} 
	\caption{Dependence of scaled large particle segregation flux on small particle concentration, $c_{s}$, and particle size ratio, $R_{S}$; curves from fits to DEM segregation velocity data. The red curve intersects each flux curve at $c_{s} = 0.5$ and the blue curve passes through each flux curve at $c_{s,peak}$, with both curves projected onto the $c_{s}-R_{S}$ plane.  The maximum flux is projected onto the $R_{S}-\Phi$ plane.  ($d_{s}=2$\,mm, $q=40$\,cm$^{2}/$s.)}\label{FluxRSC}
\end{figure}

Segregation flux curves like the solid curves shown in Figs.~\ref{FluxPoly}(b) and \ref{fig:SL2}(e-f) were generated from fits to the segregation velocity data for simulations for $1.1 \leq R_{S} \leq 3$, varied in increments of 0.1, with $d_{s} = 2$\,mm, $q = 40$\,cm$^{2}/$s and varying $d_{l}$.  For each $R_{S}$, results from three simulations with $c_{s,inlet} = 0.2,\ 0.5,$ and $0.8$ were combined to generate segregation velocity data for a wider range of $c_{s}$, compared to using only $c_{s,inlet} = 0.5$.  The large particle segregation flux is plotted vs.\ $R_{S}$ and $c_{s}$ in Fig.~\ref{FluxRSC}.  The asymmetry of the segregation flux (i.e., that the maximum flux occurs for $0<c_{s}<0.5$), is highlighted by the small particle concentration at maximum segregation flux, $c_{s,peak}$, (blue curve), and flux at $c_{s} = 0.5$ (red curve), both of which are projected onto the $c_{s}-R_{S}$ plane.  The maximum segregation flux, $\Phi_{max}$, is projected onto the $\Phi_{p,l}$-$R_{S}$ plane, and shows that $\Phi_{p,l}$ grows monotonically with $R_{S}$ for $1.1 \le R_{S} \le 2.4$ and is relatively constant for larger $R_{S}$.  This result is consistent with previous measurements of $S=d_{s}f(R_{s})$ used in Eq.~\ref{FanEquation} that show that $S$ is independent of $R_{S}$ above similar values of $R_{S}$ \citep{Schlick2015AIChE}. The increase in the segregation flux magnitude with $R_{S}$ for $1.1 \leq R_{S} \le 2.4$ and plateau for $2.4 \le R_{S} \le 3.0$ is also consistent with the observation of a maximum segregation rate at an intermediate $R_{S}$ in an annular shear cell by \citet{GollickAndDaniels}.

The results in Fig.~\ref{FluxRSC} represent 60 DEM simulations for $d_{s} = 2$\,mm and $q = 40$\,cm$^2$/s.  However, to test the effect of changing $d_{s}$, $d_{l}$, and $q$, a total of 240 DEM simulations for size bidisperse flows were performed.  Four sets of simulations were run for each $R_{S}$ with $d_{s} = 2$\,mm, $q = 40$\,cm$^{2}/$s and varying $d_{l}$; $d_{s} = 4$\,mm, $q = 40$\,cm$^{2}/$s and varying $d_{l}$; $d_{s} = 2$\,mm, $q = 20$\,cm$^{2}/$s and varying $d_{l}$; and $d_{l} = 6$\,mm, $q = 40$\,cm$^{2}/$s and varying $d_{s}$.  This results in a broad range of shear rates, $0.1 < \dot{\gamma} < 29.4$ s$^{-1}$ for $q = 20$\,cm$^{2}/$s and $0.1 < \dot{\gamma} < 45.2$ s$^{-1}$ for $q = 40$\,cm$^{2}/$s. Segregation flux results from the 180 simulations not shown in Fig.~\ref{FluxRSC} are included in the Supplementary Material and are quantitatively similar to the results in Fig.~\ref{FluxRSC}.  

\begin{figure}
	\center	
		\includegraphics[scale = 1]{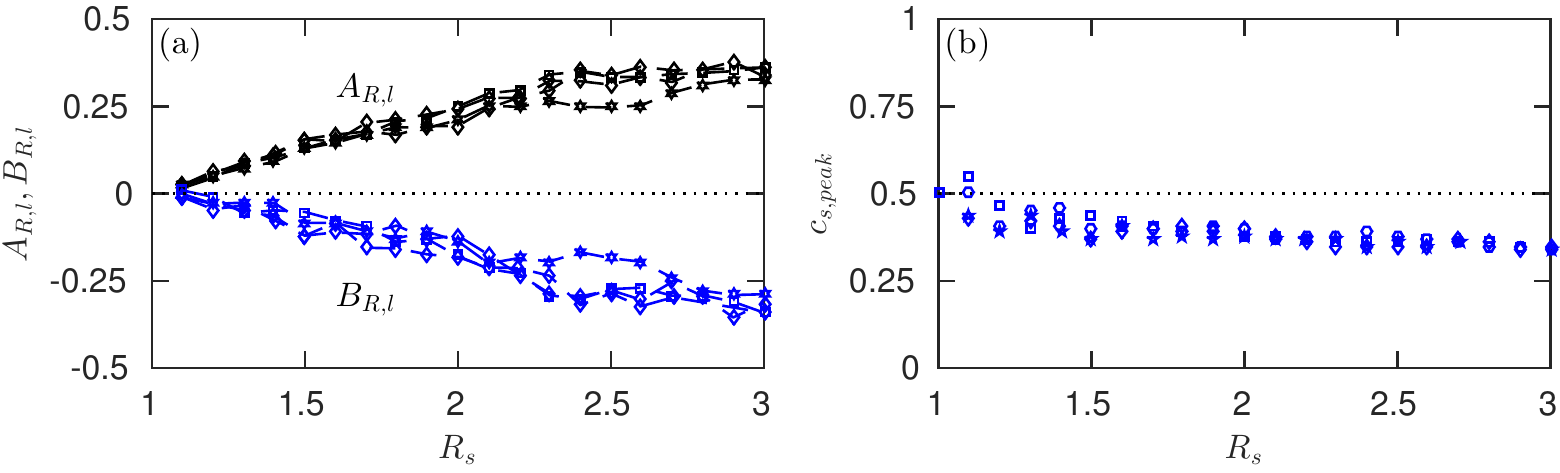}	
	\caption{(a) Quadratic segregation velocity coefficients, $A_{R,l}$ (black) and $B_{R,l}$ (blue), from Eq.~\ref{WPoly}.  (b) Small particle concentration at maximum flux, $c_{s,peak}$ for various operating conditions: $\square = d_{s}=2$\,mm, $q=40$\,cm$^{2}/$s, $d_{l}$ varies; $\circ = d_{s}=2$\,mm, $q=20$\,cm$^{2}/$s, $d_{l}$ varies; $\diamond = d_{s}=4$\,mm, $q=40$\,cm$^{2}/$s, $d_{l}$ varies; $\star = d_{l}=6$\,mm, $q=40$\,cm$^{2}/$s, $d_{s}$ varies;}\label{SCoeffs}
\end{figure}

The quadratic fit coefficients for each $R_{S}$ and the concentration of small particles at which the maximum segregation flux occurs are shown in Fig.~\ref{SCoeffs}, for all 240 DEM simulations performed.  The figure shows that the coefficients and the concentration at peak segregation flux are relatively independent of absolute particle size and flow rate for the range of particle sizes and flow rates that were examined.  The coefficients $A_{R,l}$ and $B_{R,l}$ grow from near zero at $R_{S} = 1.1$ to a plateau value near $R_{S} = 2.4$, above which they stay relatively constant.  The coefficients $A_{R,s}$ and $B_{R,s}$ for small particles can be calculated from the large particle coefficients as 
\begin{equation}
  \begin{aligned}
     A_{R,s} &= A_{R,l} + B_{R,l} \\
     B_{R,s} &= -B_{R,l}.
  \end{aligned}
\end{equation}
The small particle concentration corresponding to the maximum segregation flux, $c_{s,peak}$, shown in Fig.~\ref{SCoeffs}(b) is somewhat variable for small values of $R_{S}$, but nearly always below $0.5$.  The value for $c_{s,peak}$ gradually decreases, reaching a minimum value of $0.35$ at $R_{S} = 3$ for the range of $R_{S}$ considered.  

\section{\label{sec:SavageLun}Evaluation of Savage and Lun's kinetic sieving model}

Having confirmed that the empirically determined cubic form of the flux (Eq.~\ref{cubicFlux}) accurately describes segregation fluxes in DEM simulations of bidisperse flow, we now compare these results to the kinetic sieving segregation model proposed by \citet{Savage1988}, which is non-linear in concentration.  However, the model is challenging to implement in practice because it is difficult to unambiguously determine the various coefficients used in the model.  As a result, to the authors' knowledge, its validity has not been previously confirmed, though it is frequently cited.  Below, the model is evaluated and, for the first time, compared to DEM simulation results to evaluate its validity.  

\subsection{Savage and Lun model}

Savage and Lun's segregation model \citep{Savage1988} (hereafter referred to as the SL-model) uses a first-principles approach to predict particle segregation in moderately-sheared, gravity-driven free surface flow, specifically dense size-bidisperse mixtures of spheres flowing down a rough-bottomed chute.  Expanding on the information-entropy approach of \citet{Cooke1979}, the SL-model is based on the assumption that the probability for a small particle to fall into a void is larger than the corresponding probability for a large particle. Consequently, the unequal downward rate of void-filling induces segregation of the small particles relative to large particles, with small particles segregating below the large ones. In addition to this ``random kinetic sieving'' mechanism, the SL-model also includes a non-size-preferential term to allow the upward movement of particles, called ``squeeze-expulsion,'' which is necessary to balance the net downward flux of both species owing to kinetic sieving.  The mechanisms of random kinetic sieving and squeeze expulsion are additively combined in the SL-model to give the \textit{net} volume-averaged percolation velocity, which is analogous to the percolation velocity considered in this paper.
 
 The SL-model relates the net volume-averaged percolation velocity, $w_{p,i}$ of species $i$ to the local number ratio of small to large species, $\eta$, the shear rate, $\dot{\gamma}$, and the diameter ratio of small to large particles, $R^{-1}_{S}$:
 \begin{equation}\label{eq:1}
\begin{aligned}
 w_{p,s}&=d_{l}\dot{\gamma}\left[\frac{-1}{(1+\eta (R^{-1}_{S})^{3})}\right](w^*_{p,s}-w^*_{p,l})\\
 w_{p,l}&=d_{l}\dot{\gamma}\left[\frac{\eta (R^{-1}_{S})^{3}\,\,}{(1+\eta (R^{-1}_{S})^{3})}\right](w^*_{p,s}-w^*_{p,l}),
 \end{aligned}
  \end{equation}
  where
  \begin{equation}\label{eq:2}
  \begin{aligned}
  w^*_{p,s}&=G(\eta,R^{-1}_{S})\left[\bar{E}-E_{m}+1+\frac{(1+\eta)R^{-1}_{S}}{(1+\eta R^{-1}_{S})}\right]\\&\quad\cdot\exp\left[-\frac{(1+\eta)R^{-1}_{S}/(1+\eta R^{-1}_{S})-E_{m}}{\bar{E}-E_{m}}\right]\\
  w^*_{p,l}&=G(\eta,R^{-1}_{S})\left[\bar{E}-E_{m}+1+\frac{(1+\eta)}{(1+\eta R^{-1}_{S})}\right]\\&\quad\cdot\exp\left[-\frac{(1+\eta)/(1+\eta R^{-1}_{S})-E_{m}}{\bar{E}-E_{m}}\right],
  \end{aligned}
  \end{equation}
  and
  \begin{equation}\label{eq:3}
  G(\eta,R^{-1}_{S})=\frac{4k_{\scriptscriptstyle{LT}}^{2}(M/N)(1+\eta R^{-1}_{S})}{\pi(1+\eta)\left[\frac{(1+\eta)(1+\eta (R^{-1}_{S})^{2})}{(1+\eta R^{-1}_{S})^{2}}+\frac{\bar{E}^{2}}{k_{av}}\left(\frac{M}{N}\right)\right]}.
  \end{equation}
It is straightforward to show that the bracketed terms in Eq.~\ref{eq:1} are equivalent to $1-c_{i}$, which means that Eq.~\ref{eq:1} is equivalent to Eq.~\ref{simpleSL} with $f(c_{s},R_{S})= w^*_{p,s}-w^*_{p,l}$. 

\subsection{Fitting the SL-Model}

The challenge to using the SL-model for predicting the percolation velocity is that several parameters ($\bar{E}$, $E_{m}$, $k_{LT}$, $k_{av}$, $M/N$) must be known.  Determining these parameters directly from experiments proves difficult, as they derive from and are based solely on the assumption that distinct layers of particles in the flow exist and form a sieve-like arrangement, which, does not typically occur. For this reason, application of the full SL-model has been limited, requiring, at best, a heuristic determination of the parameter values.  However, \citet{Savage1988} proposed that $\bar{E}$, $M/N$, and $k_{av}$ could be calculated for various 2D packings of equal-sized spheres to create a physically ``consistent" set of values.  For example, for 5 equal-sized particles surrounding a void, the SL-model suggests $\bar{E}=0.701,M/N=0.6,k_{av}=0.712,k_{\scriptscriptstyle\text{LT}}=1$ 

Here, the model parameters for Eq.~\ref{eq:1} are first found through a simultaneous fit to all of the DEM segregation velocity data used to create Fig.~\ref{FluxRSC}.  The fit (MATLAB nonlinear least squares) was performed over the parameter set $\{\bar{E}, M/N, k_{av}\}$, using the lower and upper bounds proposed by \citet{Savage1988}, assuming that $k_{LT}=1$, meaning the layer thickness is equal to the mean local particle diameter.  The proposed lower and upper bounds for the parameters $\{\bar{E}, M/N, k_{av}\}$ are based on the 2D-packing of mono-disperse spheres to create a physically ``consistent" set of values \citep{Savage1988}.  The values of the SL-model parameters determined by the fit to the entire set of DEM data are $\bar{E}=0.477,M/N=0.781,k_{av}=0.466,k_{\scriptscriptstyle\text{LT}}=1$, noting that the value for $k_{av}$ is equal to the lower limit proposed by \citet{Savage1988}.  

\begin{figure}
		\centering
		\includegraphics[clip, trim=0cm 0cm 12.5cm 23cm,scale = 1]{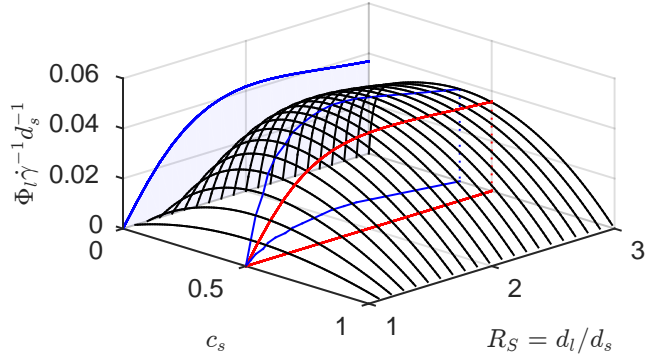}
		\caption{Normalized segregation flux from Savage and Lun's segregation model \citep{Savage1988} vs.\ particle size ratio, $R_{S}$, and small particle concentration, $c_{s}$, using parameter values obtained from a fit to DEM segregation velocity data: $\bar{E}=0.477$, $M/N=0.781$, $k_{\scriptscriptstyle{\text{av}}}=0.466$, $k_{\scriptscriptstyle{\text{LT}}}=1$.  The red curve intersects each flux curve at $c_{s} = 0.5$ and the blue curve intersects each flux curve at $c_{s,peak}$, with both curves projected onto the $c_{s}-R_{S}$ plane.  The maximum flux is projected onto the $R_{S}-\Phi$ plane.  ($d_{s}=2$\,mm, $q=40$\,cm$^{2}/$s.)}
		\label{fig:SL1}
	\end{figure}

Figure \ref{fig:SL1} shows the non-dimensional segregation flux from the SL-model vs.\ $c_{s}$ and $R_{S}$ for $1 \le R\leq 3$.  The similarity between the SL-model in Fig.~\ref{fig:SL1} and the DEM data in Fig.~\ref{FluxRSC} is remarkable.  The SL-model captures both the asymmetry in the non-dimensional segregation flux about $c_{s}=0.5$ with the maximum segregation flux in the range $0\leq c_{s,peak}<0.5$ and the dependence of non-dimensional segregation flux on $R_{S}$ which first increases with $R_{S}$ and then plateaus for higher values of $R_{S}$.  Note, however, that the magnitude of the non-dimensional flux is higher for the SL-model than the DEM simulation for small $R_{S}$ and lower for large $R_{S}$ due to using fixed parameters over all $R_{S}$ values.    

The qualitative features of the SL-model shown in Fig.~\ref{fig:SL1} persist over a relatively wide range of parameters.  We systematically explored this dependence for $\bar{E}\in\left\{0.1547,1\right\}$, $M/N\in\left\{0.5,2\right\}$, and $k_{av}\in\left\{0.466,0.765\right\}$, the parameter ranges suggested by Savage and Lun \citep{Savage1988} for 4 to 6 equal-sized particles surrounding a void. The model is relatively insensitive to the parameters in this range. 

While Fig.~\ref{fig:SL1} demonstrates that the SL-model can qualitatively capture the features of the segregation velocity and segregation flux observed in the DEM data, the match is imperfect, even though the SL-model parameters are derived from the DEM simulations.  Figure~\ref{fig:SL2}\,(d-f) compares the DEM data with both the SL-model (dashed curves) and the quadratic form for $w_{p,i}$ (Eq.~\ref{WPoly}, solid curves) in terms of both the segregation velocity and the segregation flux for three values of the size ratio, $R_{S} = 1.6,\ 2.4,$ and $2.8$.  In all cases, the quadratic form matches the data better than the SL-model.  The SL-model consistently shows a higher flux at lower values of $c_{s}$ than appears in the DEM data as well as higher magnitude of the non-dimensional flux for small $R_{S}$ and lower flux for large $R_{S}$.  To some extent, this is to be expected because the quadratic form (i.e., $A$ and $B$) is fit directly to the data for each value of $R_{S}$, whereas the parameters for the SL-model are based on the data across all $R_{S}$.  Fitting the parameter set of the SL-model to the data for a single $R_{S}$ instead of over all values of $R_{S}$ improves the fit of the SL-model to the data, shown as the dash-dot curve in Fig.~\ref{fig:SL2}(d-f), as would be expected.  The SL-model parameter values are (for $R_{S} = 1.6, 2.4,$ and $2.8$, respectively) $\bar{E}=0.487, 0.517,$ and $ 0.532$, $M/N=0.5, 0.693,$ and $ 0.812$, $k_{\scriptscriptstyle{\text{av}}}=0.466, 0.466,$ and $ 0.466$ (lower bound = $0.466$), $k_{\scriptscriptstyle{\text{LT}}}=1$ (fixed for all fits) but the fit is still not as good as the fit to the quadratic form of Eq.~\ref{WPoly}.  Nevertheless, one can conclude that the SL-model, while effective in predicting the qualitative dependence of the segregation flux on small particle concentration and size ratio, does not accurately predict the segregation velocity or flux.  This suggests that the kinetic sieving and squeeze expulsion mechanisms incorporated in the SL-model are correct to first order, but oversimplify the actual physics at play in practical segregation situations.  

\section{\label{sec:Density} Density segregation}

It is natural to also consider density-driven segregation given  that the segregation velocity in density-driven segregation of particles of the same size can be predicted using Eq.~\ref{FanEquation}, where $S$ depends on the particle density ratio instead of the particle size ratio \citep{XiaoDensity}.  Here we explore how the segregation flux, $\Phi_{i}$, varies with the density ratio, $R_{D}$, and the concentration of heavy particles, $c_{H}$.  

Using the same DEM methodology as with size bidisperse granular flows in the geometry shown in Fig.~\ref{Heap}, density bidisperse flows of $d=3$\,mm spherical particles were evaluated to characterize the concentration dependence of the segregation velocity.  Results from 30 DEM simulations were used to evaluate the segregation velocity for $1 \leq R_{D} \leq 10$ in increments of $1$.  The density of light particles, $\rho_{L}$ was fixed at 2500\,kg/m$^{3}$ and the density of heavy particles, $\rho_{H}$, was varied from 2500 to 25000\,kg/m$^{3}$; the feed rate was $40$\,cm$^{2}$/s.  The particle diameters were uniformly distributed between $2.7$\,mm and $3.3$\,mm to reduce particle ordering.  DEM simulation results were processed in the same manner as for size bidisperse simulations to calculate the segregation velocity, species concentration, and shear rate data from three simulations with differing values of $c_{H,inlet}$ to calculate the segregation velocity over the full range of mixture concentrations for each $R_{D}$ value.

\begin{figure}
	\center	
		\includegraphics[scale = 1]{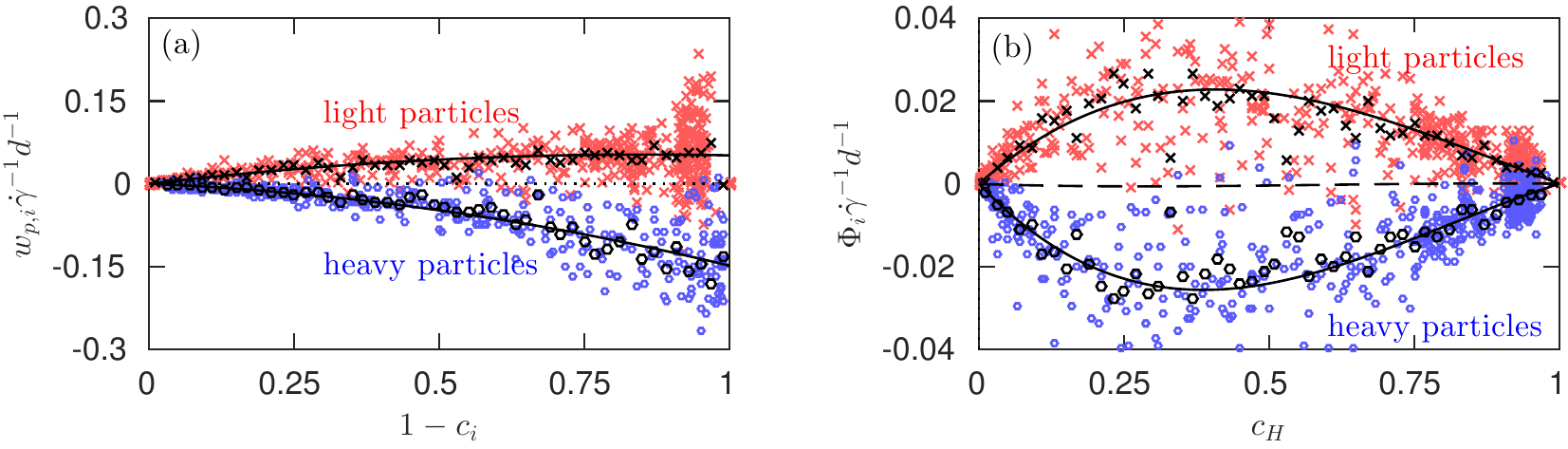}	
	\caption{Density-driven segregation velocity vs.\ $1-c_{i}$ and segregation flux vs.\ local heavy particle concentration, $c_{H}$, for $R_{D} = 6$, $\rho_{L} = 2500$\,kg/m$^3$, $d=3$\,mm, and $q=40$\,cm$^2/$s.  (a) Segregation velocity from simulations with particle inlet ratios $c_{H,inlet}=0.2$, $0.5$, and $0.8$ for light (red x) and heavy (blue o) particles and black data points are averaged over 0.02 wide increments of $1-c_{i}$.  Solid (--) curve is a quadratic fit using all data in plot.  (b) Segregation flux with solid (--) flux curves from quadratic segregation fit in (a).  Dashed (- -) line is the sum of the large and small particle fluxes determined from the segregation velocity fits.}	\label{Fig:DensityVel}
\end{figure}

An example of the resulting segregation data is shown in Fig.~\ref{Fig:DensityVel} for $R_{D}=6$.  Figure~\ref{Fig:DensityVel}(a) shows the segregation velocity data vs.\ $1-c_{i}$ and the associated quadratic fit.  Figure~\ref{Fig:DensityVel}(b) shows the same data along with the segregation flux curves from the fit to the segregation velocity data.  Although there is more scatter in the data than for the case of size-driven segregation, the asymmetry in the density-driven segregation flux still occurs with $c_{H,peak} < 0.5$, as highlighted by the concentration-averaged data.  

\begin{figure}
	\center	
		\includegraphics[clip, trim=0cm 0cm 12.5cm 23cm,scale = 0.99]{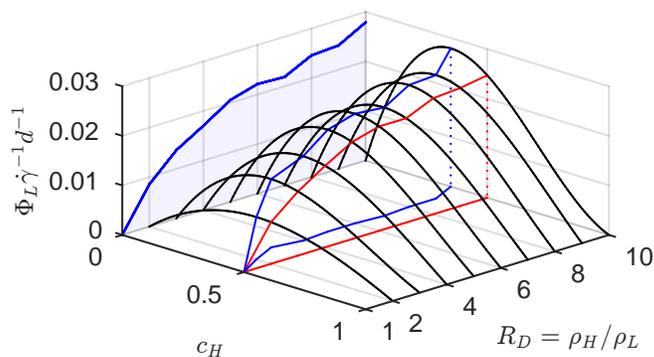} 
	\caption{Variation of the segregation flux from DEM simulations for density bidisperse flows.  Black curves show heavy particle flux from the fit to segregation velocity for each $R_{D}$ value evaluated; red curve passes through each flux curve at $c_{H} = 0.5$; blue curve passes through each flux curve at $c_{H,peak}$.  The maximum flux is projected onto the $R_{D}-\Phi$ plane.  $d=3$\,mm, $q=40$\,cm$^{2}$/s, $\rho_{L} = 2500$\,kg/m$^{3}$.}\label{densityFlux}
\end{figure}

The segregation flux from the segregation velocity fits to the DEM data for a range of $R_{D}$ is shown in Fig.~\ref{densityFlux}, again this data spans a wide range of shear rates, $0.1 < \dot{\gamma} < 38.7$ s$^{-1}$.  The results for density bidisperse segregation are qualitatively similar to those for size bidisperse segregation.  However the maximum segregation flux continues to increase with $R_{D}$ for $1 \leq R_{D} \leq 10$ rather than plateauing as in the size bidisperse case.  The segregation flux magnitude is significantly smaller for density segregation than for size segregation at equal values of $R_{D}$ and $R_{S}$, consistent with a previous review of segregation \citep{Williams1976}.  

The overall similarity between the dependence of segregation flux on $c_{s}$ and $R_{S}$ for size bidisperse mixtures and on $c_{H}$ and $R_{D}$ for density bidisperse mixtures is striking.  The dependence of density segregation on concentration is the same as for size segregation (Eq.~\ref{WPoly}) and has a qualitatively similar dependence for the flux on concentration and density ratio.  Thus the mobility of heavy particles in density bidisperse flows appears to be similar to that of small particles in size bidisperse flows, although the mechanism of size-driven preferential kinetic sieving \citep{Savage1988} should not occur in density-driven segregation of equal diameter particles.   

A possible explanation for the similarity between size- and density-driven segregation is that the mechanisms in both cases are consistent with kinetic sieving and squeeze expulsion \citep{Savage1988}.  Regardless of the size or density of a particle, the only way for it to move upward is by direct contact with particles below applying contact forces to push it upward into a void above (squeeze expulsion).  However, falling downward requires only a void below the particle.  In size bidisperse systems, small particles can fall by gravity into a smaller void than large particles.  Since larger voids occur less frequently, downward motion of small particles is preferred.  For density-bidisperse systems, a particle needs only gravity to fall into a void below it, regardless of its density.  However, to move upward, it needs both a void above it and, simultaneously, particles below providing adequate contact forces to push it upward.  We speculate that because a lighter particle is easier to push upward than a heavy particle there is a preference for upward motion of light particles, while there is an equal probability of downward motion for both light and heavy particles into a void below \citep{Jain2005}.  This preference in density bidisperse systems for light particles to preferentially be forced upward is similar to that suggested by \citet{Savage1988} for size bidisperse systems where small particles preferentially fall into voids while both large and small particles are equally likely to be pushed upward by ''squeeze expulsion."  Unfortunately, the segregation velocity results do not provide direct insight into the mechanism for density driven segregation so the above explanation is speculative.  Nonetheless, the similarities between density-driven segregation and size-driven segregation are remarkable and worthy of further investigation.  

\section{\label{sec:Conclusion}Conclusion}
Evaluating the dependence of the segregation velocity, $w_{p,i}$, on local mixture concentration for a range of size and density ratios in flowing granular material leads to several conclusions.  First, our results confirm that like size-bidisperse granular materials in a quasi-static shear cell \citep{VaartAndGray} or in an annular shear cell \citep{GollickAndDaniels}, dense surface flows of particles segregate more quickly at low concentrations of small particles in gravity-driven free surface flows.  In other words, small particles among many large particles segregate more quickly than large particles among many small particles, as evidenced by the maximum segregation flux occurring for $0.35 \leq c_{s} \leq 0.5$ over a wide range of size ratios (see Fig.~\ref{FluxRSC}). 

In a size bidisperse flow, the segregation flux increases with $R_{S}$ up to $R_{S} \approx 2.4$, above which it plateaus for $2.4 \leq R_{S} \leq 3$.  Using parameters extracted from DEM simulations, we implemented the SL-model, which, to our knowledge, is the first time this has been done, in order to validate it against segregation data.  The model \citep{Savage1988} is qualitatively similar to the DEM simulation data (compare Figs.~\ref{FluxRSC} and \ref{fig:SL1}), supporting the plausibility of the kinetic sieving mechanism and the SL-model.  The SL-model also predicts the decrease in small particle concentration at which the peak flux occurs with increasing size ratio.  However, the SL-model only matches the DEM simulation results qualitatively.  

Surprisingly, density bidisperse flows have a similar dependence on local mixture concentration and density ratio to that predicted by the SL-model and observed in size-driven segregation.  It is clear that the kinetic sieving and squeeze expulsion mechanisms must be different for density-driven segregation than for size-driven segregation, though it might be better to think in terms of a more simplistic explanation: small particles fall downward into voids more easily than large particles simply because smaller voids are more common; similarly, lighter particles are more likely to be pushed up into voids above them than heavy particles, possibly because less force is necessary to do so.

In past work an advection-diffusion-segregation continuum approach \citep{Dolgunin1995, Dolgunin1998, Gray2006, Thornton2006, Tunuguntla2014, Deng2018, Sam2018, Bridgwater1985} has been proposed to model segregation in dense surface flows of granular materials.  Along these lines, we have used Eq.~\ref{FanEquation} with a linear dependence of the segregation on concentration to model segregation for a range of size bi- and polydisperse flows and density bidisperse flows \citep{Schlick2015AIChE, SchlickPoly, XiaoDensity, Schlick2015, Fan2014} and shown that the results match DEM results quite well.  However, in this paper we have demonstrated that $w_{p,i}$ is better fit by a quadratic dependence on concentration than a linear fit, resulting in a dependence for the segregation flux (Eq.~\ref{cubicFlux}) identical to that proposed previously by \citet{GajjarAndGray}.  Hence, one might question if using a concentration asymmetric model (Eq.~\ref{WPoly}) in the advection-diffusion-segregation model would work better.  To address this question we compared results of our advection-diffusion-segregation model using both Eq.~\ref{FanEquation} and Eq.~\ref{WPoly} to DEM simulation results.  Using Eq.~\ref{WPoly} in the model matches DEM simulation results only slightly better than using Eq.~\ref{FanEquation}.  The differences, although relatively small, are most prominent as the concentration deviates from $c_{s} = 0.5$.  Further details are provided in the supplementary material.

While several questions regarding the segregation velocity and flux have been answered in this study, further research is needed.  The reason for the loss of dependence of segregation on the particle size ratio for $R_{S} > 2.4$ is unclear.  Also of interest is an explanation for why both size and density driven segregation produce similar segregation velocity and segregation flux relations even through the mechanisms would seemingly be quite different.  An additional challenge is to connect the segregation velocity with the driving forces on the segregating particle for a particle at the dilute concentration limit \citep{Guillard2016}.    Additional work is also necessary to consider segregation outside the range of bidisperse mm-sized particles with small size ratios considered here. Of particular interest in industrial and geophysical flows is segregation of polydisperse particles having a continuously varying range of size ratios that can exceed two orders of magnitude or more \citep{Marks2015}.

\section*{\label{sec:Acknoledgements}Acknowledgments}
Funded by The Dow Chemical Company and NSF Grant No. CBET-1511450.

\bibliography{AsymBib}

\clearpage
%\newpage
\section*{Supplementary Material}
\clearpage
%\newpage

\section*{\label{sec:SupFluxCurves}Segregation flux dependence on size ratio}

\setcounter{figure}{0}
\setcounter{page}{1}
\setcounter{equation}{0}

\begin{figure*}
% From Paul: Include a color bar for concentration.
	\center	
		\includegraphics[scale = .9]{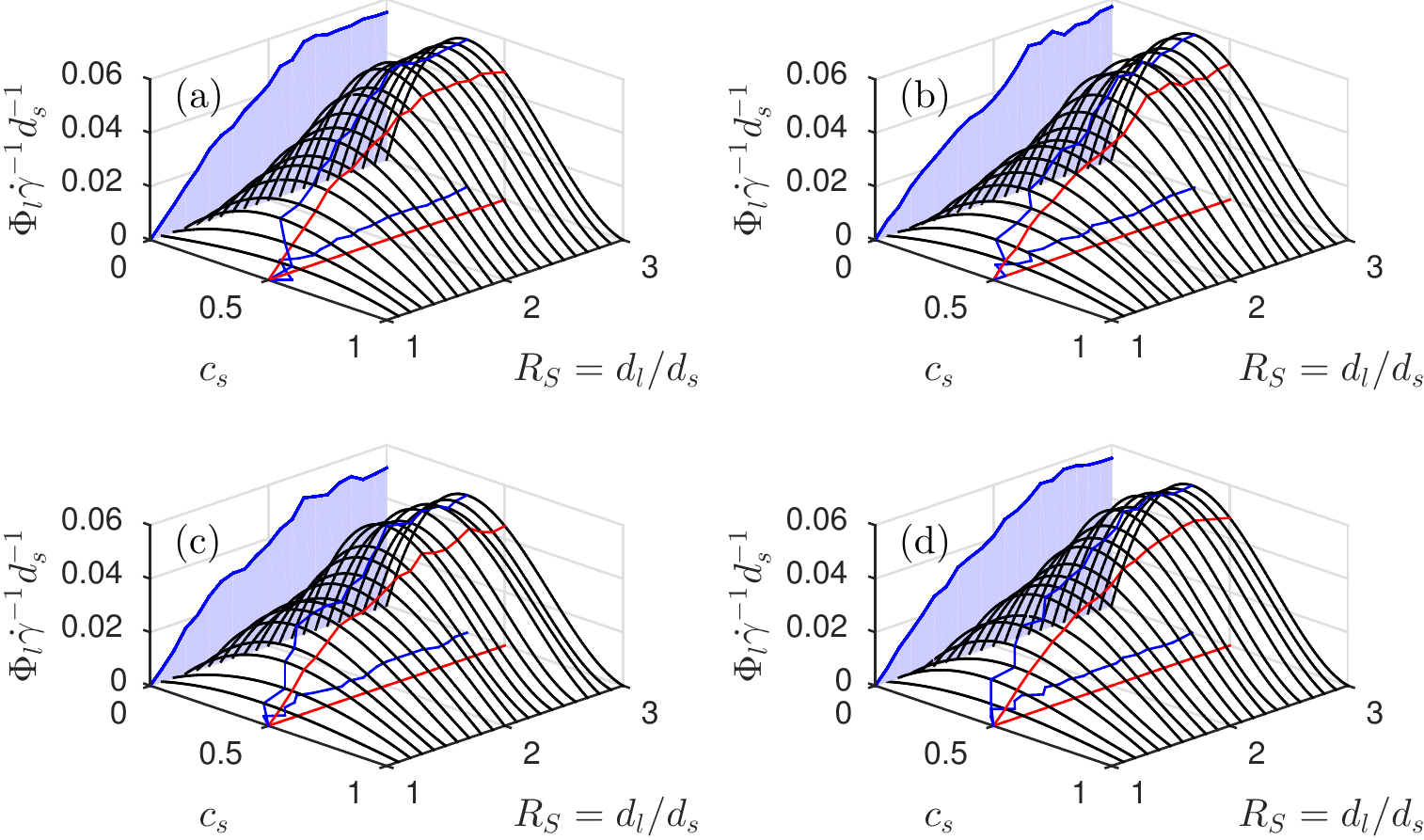} 	
	\caption{Large particle segregation flux curves from fits to DEM segregation velocity data at various particle size ratio, $R_{S}$, values, showing the variation of large particle segregation flux with $R_{S}$ and small particle concentration, $c_{s}$. The gray (red online) curves intersect each flux curve at $c_{s} = 0.5$ and the black (blue online) curves intersect each flux curve at $c_{s,peak}$, with corresponding projections onto the $c_{s}-R_{S}$ plane.  The maximum flux is projected onto the $R_{S}-\Phi$ plane.  (a) $d_{s}=2$\,mm, $q=40$\,cm$^{2}/$s, $d_{l}$ varies; (b) $d_{s}=2$\,mm, $q=20$\,cm$^{2}/$s, $d_{l}$ varies; (c) $d_{s}=4$\,mm, $q=40$\,cm$^{2}/$s, $d_{l}$ varies; (d) $d_{l}=6$\,mm, $q=40$\,cm$^{2}/$s, $d_{s}$ varies;}\label{fluxCompare}
\end{figure*}

The segregation flux curves plotted against the size ratio, $R_{S}$, and the concentration of small particle, $c_{s}$, for all 240 DEM simulations are shown in Fig.~\ref{fluxCompare}.  The 240 simulations were organized into four sets of data for different $q$, $d_{s}$, and $d_{l}$ values, and show that the segregation flux depends on the local concentration, shear rate ($\dot{\gamma}$ in the segregation velocity) and particle size ratio $R_{S}$ and is independent of flow rate, $q$, and the absolute particle size for mm-sized particles.

\section*{\label{sec:Model}Advection-diffusion-segregation modeling}

Advection-diffusion-segregation models have been used in various forms to predict segregation in flowing granular materials \citep{Dolgunin1995, Dolgunin1998, Fan2014, Schlick2015AIChE, XiaoDensity, Gray2006, Thornton2006, Tunuguntla2014} ever since the approach was first proposed by \citet{Bridgwater1985} over thirty years ago.  We have successfully applied the model using a segregation velocity linear with concentration (Eq.~\ref{FanEquation}) for a variety of situations including bi- and polydisperse size segregation in bounded heap flow \citep{Schlick2015AIChE, SchlickPoly, Fan2014}, bidisperse segregation in tumbler flow \citep{Schlick2015}, tri-disperse size segregation in chute flow \citep{Deng2018}, bidisperse density segregation in bounded heap flow \citep{XiaoDensity}, and even segregation of rod-like particles in bounded heap flow \citep{Sam2018}.  Here we compare the predictions of the model using a segregation velocity linearly dependent on $c_{s}$ (Eq.~\ref{FanEquation}) and a segregation velocity quadratically dependent on $c_{s}$ (Eq.~\ref{WPoly}) to results from DEM simulations.    

The segregation velocity can be used to predict concentration fields in various flow geometries and for various initial conditions using an advection-diffusion-segregation model of the form
\begin{equation}
 \label{ADSmodel} 
 \dfrac{\partial c_{i}}{\partial t} +  \nabla \cdot \mathbf{u}c_{i} + \dfrac{\partial}{\partial z}w_{p,i}c_{i} =  \nabla \cdot (D\nabla c_{i}),
\end{equation}
which has been shown to match experimental and DEM results well \citep{Fan2014, SchlickPoly, XiaoDensity, Schlick2015, Schlick2015AIChE} using Eq.~\ref{FanEquation} for the segregation velocity in the segregation term $^{\partial}/_{\partial z}(w_{p,i}c_{i})$.  The advective term $\nabla \cdot (\mathbf{u}c_{i})$, where $\mathbf{u}$ is velocity vector, and the diffusion term $\nabla \cdot (D\nabla c_{i})$, where $D$ is the diffusion coefficient, both depend on the physical characteristics of the flow which can be determined from DEM results, theory (for simple geometries), or experiments.    

Here we consider steady flow in a bounded heap geometry as shown in Fig.~\ref{Heap}.  The segregation velocity coefficient, $S$, for the linear segregation velocity is $S = d_{s}C\log (R)$ with $C$ = 0.26 \citep{Schlick2015AIChE}, while the segregation velocity determined by using Eq.~\ref{WPoly} uses $A = 0.29$ and $B = -0.22$ as shown in Fig.~\ref{FluxPoly}(b) for $R_{S} = 2.2$.

\begin{figure*}
% From Paul: Include a color bar for concentration.
	\center	
	\hspace*{\fill}
		\raisebox{-\height}{\includegraphics[scale = .9]{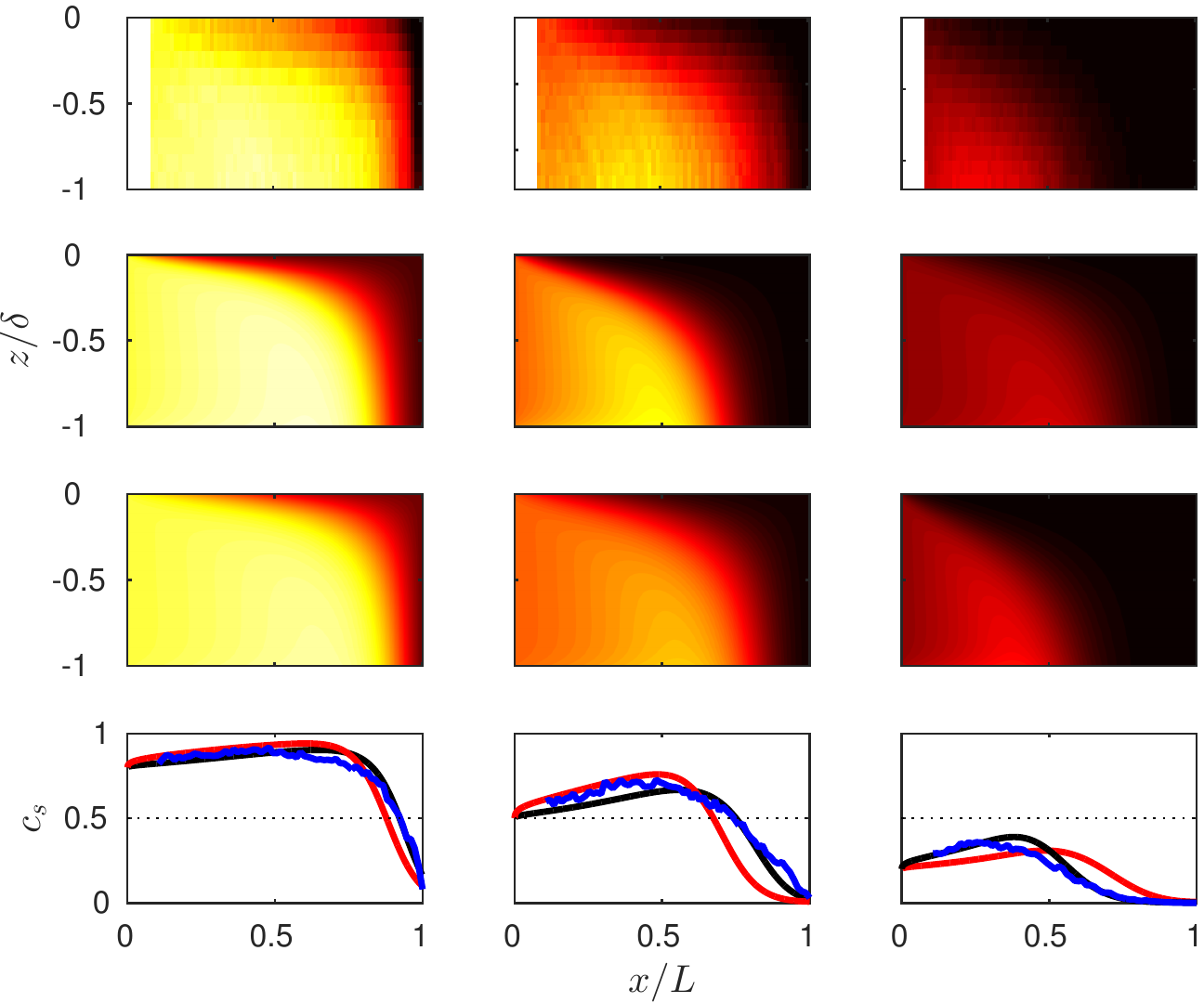} }
		\hfill
		\raisebox{-\height}{\includegraphics[scale = 0.675, trim={4.25in 0in 0.25in 0in}, clip]{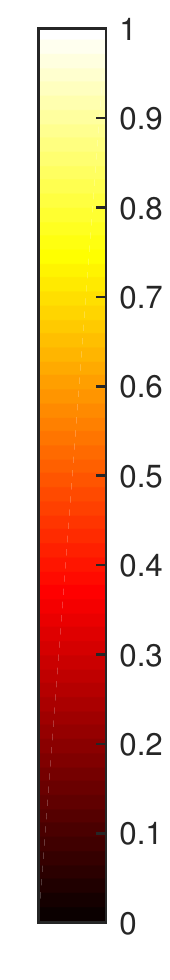}}
%		\raisebox{-\height}{\includegraphics[scale = 0.675, trim={4.25in 0in 0in 0in}, clip]{Continuum_colorbar}}		
		\raisebox{-1.2in}{\includegraphics[scale = 1.25, trim={5.0in 1.9in 0in 1.8in}, clip, angle = -90]{Continuum_colorbar-eps-converted-to}}
		\hspace*{\fill}
	\caption{Small particle concentration in the flowing layer for $c_{s,inlet}=0.2$ (left column), $0.5$ (middle column), and $0.8$ (right column) for DEM simulations (top row) and from advection-diffusion-segregation model with linear ($2^{nd}$ row) and quardratic ($3^{rd}$ row) segregation velocity. (Bottom row) Comparison of small particle deposition onto the heap between DEM simulation (blue line), model with linear (red line), and quadratic velocity (black line) dependence of percolation velocity on concentration.}\label{modelResults}
\end{figure*}

Figure \ref{modelResults} compares results from DEM simulations to the predictions of the continuum model for three different inlet conditions, $c_{s,inlet}=0.2$ (left), $0.5$ (center), and $0.8$ (right).  The top row is DEM simulation data for $c_{s}$ in the flowing layer.  The second row shows $c_{s}$ in the flowing layer from the continuum model with a linear segregation velocity relationship (Eq.~\ref{FanEquation}) and the third row shows results for quadratic segregation velocity (Eq.~\ref{WPoly}).  In all cases, the top and bottom of each image represents the surface and bottom of the flowing layer, respectively, the left side represents the inlet, and the right side represents the downstream end wall.  The bottom row shows the small particle concentration deposited onto the heap for DEM simulation and the model.  

The top three rows of the figure show little qualitative difference for the small particle concentration in the flowing layer.  The difference between the linear and quadratic segregation velocity models is more evident for the concentration of small particles deposited on the heap, shown in the bottom row of Fig.~\ref{modelResults}.  Using the quadratic form for the segregation velocity reduces the difference between the DEM data and the model predictions. The improvement is most prominent at $c_{s,inlet} = 0.2$ and $c_{s,inlet} = 0.8$.  However, the improvement is relatively small, particularly given the other assumptions of the model.

\end{document}